# Parallel transport and layer-resolved thermodynamic measurements in twisted bilayer graphene

G. Piccinini,[1,2] V. Mišeikis,[2,3] K. Watanabe,[4] T. Taniguchi,[5] C. Coletti[2,3,*], S. Pezzini[6,**]

[1]NEST, Scuola Normale Superiore, Piazza San Silvestro 12, 56127 Pisa, Italy
[2]Center for Nanotechnology Innovation @NEST, Istituto Italiano di Tecnologia, Piazza San Silvestro 12, 56127 Pisa, Italy
[3]Graphene Labs, Istituto Italiano di Tecnologia, Via Morego 30, 16163 Genova, Italy
[4]Research Center for Functional Materials, National Institute for Materials Science, 1-1 Namiki, Tsukuba, 305-0044, Japan
[5]International Center for Materials Nanoarchitectonics, National Institute for Materials Science, 1-1 Namiki, Tsukuba, 305-0044, Japan
[6]NEST, Istituto Nanoscienze-CNR and Scuola Normale Superiore, Piazza San Silvestro 12, Pisa 56127, Italy

**Abstract:** We employ dual-gated 30°-twisted bilayer graphene to demonstrate simultaneous ultra-high mobility and conductivity (up to 40 mS at room temperature), unattainable in a single-layer of graphene. We find quantitative agreement with a simple phenomenology of parallel conduction between two pristine graphene sheets, with a gate-controlled carrier distribution. Based on the parallel transport mechanism, we then introduce a method for *in situ* measurements of the chemical potential of the two layers. This twist-enabled approach, neither requiring a dielectric spacer, nor separate contacting, has the potential to greatly simplify the measurement of thermodynamic quantities in graphene-based systems of high current interest.

**Main text:** Due to its remarkable charge carrier mobility, graphene is a promising candidate for high-performance devices in applicative fields such as high-frequency electronics and optoelectronics [1–3]. In the case of graphene multilayers, the electrical transport properties strongly depend on the number of layers, as well as on the twist angle between them [4]. In particular, turbostratic multilayer graphene was shown to (potentially) harbor not only a mobility as high as single-layer graphene (SLG), but also larger conductivity ($\sigma$) [5–9]. This is possible since a finite (typically uncontrolled) twist angle between the layers preserves the gapless linear band dispersion [10], while charge redistribution keeps a relatively low carrier density ($n$) in the different layers [11]. Moreover, it was suggested that the lowest graphene sheet can efficiently screen the others from detrimental substrate-induced potential fluctuations [12]. However, thus far, experiments addressing the carrier mobility of rotated graphene multilayers have been limited to samples with considerable extrinsic disorder (preventing accurate testing of the transport performance) and a single gate electrode (lacking control on the interlayer charge distribution). We present transport measurements on a dual-gated 30°-twisted bilayer graphene (30TBG) device encapsulated in hexagonal Boron Nitride (hBN). Thanks to low-energy interlayer decoupling and high device quality, we show that 30TBG replicates the transport properties of two pristine SLG sheets conducting in parallel, including ultra-high mobility at large carrier concentration.

Furthermore, we show that dual-gated 30TBG enables *in situ* – or, better said, *in device* – thermodynamic measurements on the individual graphene layers. Measuring the chemical potential ($\mu$) as a function of experimental knobs such as *n* or an external magnetic field (*B*) can greatly contribute to the understanding of novel two-dimensional systems. To this end, probes such as scanning single-electron transistors have been widely employed on graphene [13,14]. Experiments based on capacitance spectroscopy can be performed using either metallic [15,16] or graphitic gates [17]. Kim *et al.* introduced a double-SLG configuration, allowing reciprocal measurements of $\mu(n,B)$ via dc electrical transport [18]. This technique, however, requires a dielectric spacer (typically few-nanometer thick hBN) to ensure capacitive coupling between separately-contacted graphene sheets [19,20,21]. Here, we exploit the effective electronic decoupling of large-angle TBG to obtain equivalent information via dc electrical transport in a standard dual-gated device. In particular, by keeping one of the layers charge-neutral, it is possible to probe $\mu$ in the other one with a resolution in the meV range (comparable to hBN-spaced structures [20]), as demonstrated by measurements of the SLG Landau level (LL) energies at *B* = 250 mT.

30TBG is synthesized via chemical vapor deposition (CVD) on Cu, encapsulated in hBN and processed into a dual-gated device (see sketch in Fig. 1(a) and Supplemental Material [22] for details), using the methods developed in Ref.[23]. The selective growth of 30TBG on Cu is favored by the interaction between step edges on the Cu surface and the graphene layers, which triggers the orientation of straight (either zigzag or armchair) graphene edges along the step direction [24]. Transfer of 30TBG to SiO$_2$/Si and hBN-mediated pickup are performed while always keeping *T* < 170 °C to avoid possible relaxation of the twist angle. The stability of the 30°-rotated configuration upon hBN encapsulation was confirmed by transmission electron microscope selected area electron diffraction, as reported in Ref.[23]. The assembly technique by Purdie *et al.* [25] is used to promote interface cleaning and obtain high-quality electronic transport.

The top- and back-gate voltages ($V_{tg}$ and $V_{bg}$) couple to TBG via hBN (32 nm thickness, determined by atomic force microscopy) and in-series SiO$_2$-hBN (285-40 nm thickness), respectively, resulting in the capacitance-per-unit-area $C_{tg}$ = 8.3×10$^{-8}$ F/cm$^2$ and $C_{bg}$ = 9.8×10$^{-9}$ F/cm$^2$. In addition, it is well established that a considerable interlayer capacitance $C_{gg}$ = 7.5×10$^{-6}$ F/cm$^2$ must be considered for large-angle TBG [26,27]. Data in Figure 1(b) show the longitudinal resistivity of 30TBG measured at room *T*, as a function of $V_{tg}$ and $V_{bg}$. The high-resistivity diagonal ($\rho$ up to ~400 Ω) corresponds to the *global* charge neutrality point (CNP) of the sample, while $\rho$ as low as 25 Ω is measured at large carrier concentration. The fact that the global CNP crosses exactly $V_{tg}$ = $V_{bg}$ = 0 V indicates negligible residual doping in the device. As shown by the curves in Fig. 1(c) inset, the resistivity peak becomes wider and shallower upon unbalancing $V_{tg}$ and $V_{bg}$, i.e., by increasing the so-called displacement electric field (*D*). This is caused by a splitting of the CNPs for the two layers [27], although not resolvable at room *T* due to thermal broadening. To extract quantitative information from the transport data, it is necessary to have knowledge of the *individual* carrier density in the layers ($n_{upper}$ and $n_{lower}$), which can be obtained by electrostatic modelling of the gated TBG system [26,27]; complete details on our procedure are provided in Supplemental Material [22]. Figure 1(c) shows the calculated $n_{upper}$ and $n_{lower}$ as a function of $V_{tg}$, along the straight trajectories indicated by lines in Figure 1(b), which correspond to fixed differences in the gate potentials (weighted with their respective gate capacitance). We note that such trajectories do not coincide with constant values of displacement field (as defined in the electrostatic model in Supplemental Material [22]), unless at *D* = 0 (black line in Figure 1b). While at *D* = 0 the layers are charge-balanced (black line, $n_{upper}$ = $n_{lower}$), increasingly separated and nonlinear $V_{tg}$-dependences are observed otherwise. Complete color plots of the $V_{tg}$-$V_{bg}$ dependence of $n_{upper}$ and $n_{lower}$ are shown in Fig. S1(a) and S1(b) in Supplemental Material [22]. Based on such relations, we can calculate the total carrier density in the system ($n_{tot}$ = $n_{upper}$ + $n_{lower}$) and use the standard Drude model to convert the experimental $\rho$ data into the mobility $\frac{1}{e|n_{tot}|\rho}$ shown in Fig. 1(d). At $n_{tot}$ < $10^{12}$ cm$^{-2}$, we observe a carrier mobility as high as 1.9×10$^5$ cm$^2$V$^{-1}$s$^{-1}$, and a weak dependence on the interlayer charge configuration, ascribable to the *D*-induced broadening of the resistivity peak. However, such difference becomes irrelevant at $n_{tot}$ > $10^{12}$ cm$^{-2}$, where the four experimental curves collapse on each other. Here, the room *T* mobility of 30TBG strikingly surpasses the intrinsic phonon-limited behavior of SLG (blue dotted line) [28,29], although falling below the theoretical limit for two parallel-conducting SLG for $n_{tot}$ < 4×10$^{12}$ cm$^{-2}$ (blue solid line, calculated assuming $n_{upper}$ = $n_{lower}$ and a density-independent resistivity of 52 Ω in each layer [28]). This fact can be reasonably expected based on experiments on hBN-encapsulated SLG, where a room *T* mobility below the phonon limit is observed at relatively low carrier density (*n* < 2×10$^{12}$ cm$^{-2}$) [30]. For $n_{tot}$ > 4×10$^{12}$ cm$^{-2}$, 30TBG perfectly mimics two phonon-limited SLG that simply conduct in parallel, leading to an unprecedented room *T* mobility at large carrier density (~5×10$^4$ cm$^2$V$^{-1}$s$^{-1}$ at $n_{tot}$ ~ 5×10$^{12}$ cm$^{-2}$, corresponding to $\sigma$ ~ 40 mS). Such combination might have important applicative implications for high-speed electronics [7,31–33], integrated optoelectronics [34,35], power conversion efficiency in solar cells [36,37], and sensing [38,39].

In Fig. 2, we further investigate the parallel transport mechanism in TBG at low temperature ($T$ = 4.2 K). We again consider several trajectories at fixed gate difference (lines in Fig. 2(a) inset). However, to avoid a small resistive feature attributable to a contact malfunctioning at $V_{bg}$ ~ 0 V ($V_{tg}$-independent feature in Fig. 2(a) inset, see also Fig. S2 in Supplemental Material [22]), we consider half of the $\rho$ curves from the lower left quadrant (for $n_{tot}$ < 0), half from the upper right one (for $n_{tot}$ > 0). At low displacement field, the resistivity shows a sharp peak centered at $n_{tot}$ = 0, confirming the high quality of the CVD-grown crystals and the low disorder in the hBN-encapsulated device (Fig. 2(a), see also discussion in Supplemental Material [22]). With increasing $D$, strong broadening and attenuation (see horizontal markers in Fig. 2(a)) of the peak are observed. Fig. 2(b) shows a zoom-in for $\rho \leq 60\ \Omega$, which allows evaluating the behavior at large $n_{tot}$. We find that the 30TBG resistivity perfectly reproduces that of two parallel-conducting graphene layers, in which the carrier mean free path ($l_{mfp}$) is set by the width of the device channel 2.2 µm, and the conductivity in each SLG is given by $\sigma_{layer} = 2\frac{e^2}{h} l_{mfp} \sqrt{\pi n_{layer}}$. This indicates predominant scattering at the device boundaries, consistent with results on hBN-encapsulated SLG [30]. The large-$n_{tot}$ limit is independent of the interlayer carrier distribution, consistent with the observations at room $T$. However, we find considerable effects on the transport properties at low carrier density. In SLG, the $n^*$ parameter – the characteristic density at which log($\sigma$) switches from saturated (electron-hole puddles dominated) to linear-in-log($n$) (single-carrier type) – is determined by long-range disorder and correlates with the inverse of the carrier mobility [40]. In TBG, we find that large values of $n_{tot}^*$ can instead be determined solely by the screening of the applied electric field, and do not necessarily correspond to an increased disorder level. The vertical lines in Fig. 2(c) indicate the $n_{tot}^*$ values obtained from parallel transport simulations along the usual trajectories (assuming $l_{mfp}$ = 2.2 µm), while the experimental points show the measured peak mobility (maximum values in Fig. 2(c), inset) as a function of the extracted $n_{tot}^*$. A peak mobility ~4×10$^5$ cm$^2$V$^{-1}$s$^{-1}$ is observed at $n_{tot}^*$ > 10$^{11}$ cm$^{-2}$ (orange diamond). For comparison (blue dotted line in Fig. 2(c)), two parallel-conducting SLG with a disorder-determined $n_{tot}^*$ = 10$^{11}$ cm$^{-2}$ (i.e., 5×10$^{10}$ cm$^{-2}$ per layer) are expected to show a mobility ~8×10$^4$ cm$^2$V$^{-1}$s$^{-1}$ according to the 1/$n^*$ dependence for SLG reported in Ref.[40]. Such discrepancy should be taken into careful account, especially in experiments on TBG employing a single gate electrode. The quantitative agreement, both for $\rho$ and $n_{tot}^*$, with simple parallel transport indicates that the interlayer conductivity is negligible with respect to that along the constituent layers [41]. At 30° twisting, the suppression of interlayer transport might be further favored by the incommensurate stacking configuration [42].

In the following we show the possibility of employing one of the two graphene sheets (*the probe*) to sense the chemical potential and the carrier concentration in the other one (*the target*). The key strategy consists in the individuation and tracking of the probe CNP in the parallel transport measurements as a function of the gate potentials. In Fig. 3(a) we plot the low $T$ conductivity as a function of $V_{tg}$ relative to the global CNP, here defined as $\Delta V_{tg} = V_{tg} + V_{bg} \times C_{bg}/C_{tg}$, and $V_{bg}$. In this zoomed plot, we clearly observe a nonlinear separation between the upper and lower CNPs, resulting from the interlayer capacitive coupling [27]. The CNPs form an hourglass-like shape, separating regions with equal (*e-e* and *h-h*, high $\sigma$) and opposite carrier sign (*e-h* and *h*-e, low $\sigma$) on the two layers. The *e-h* coexistence, expected from calculations of $n_{upper}$ and $n_{lower}$ (see Fig. S1(c) and S1(d) in Supplemental Material [22]), is demonstrated by magnetotransport measurements at $V_{bg}$ = -40 V shown in Fig. S3 in Supplemental Material [22]. We note that the lower conductivity observed in the *e-h* configuration is due to the smaller total carrier density and its reduced gate-dependence (see Fig. S4 in Supplemental Material [22]). The position of the probe CNP can be accurately determined by considering log($\sigma$) as a function of log($\Delta V_{tg}$) along horizontal cuts from Fig. 3(a), and applying a procedure similar to that commonly used to extract $n^*$ in SLG. As shown in Fig. 3(b), crossing one CNP determines a saturation of log($\sigma$), due to the transition from *e-e* to *h-e* (*h-h* to *h-e*) configurations. At the probe CNP, it is possible to calculate both $n$ and $\mu$ in the target layer, using simple relations [20]: $n = C_{bg}V_{bg}/e + (C_{bg}+C_{gg})C_{tg}V_{tg}/(eC_{gg})$ and $\mu = -eC_{tg}V_{tg}/C_{gg}$ when the upper layer probes the lower one, $n = C_{tg}V_{tg}/e + (C_{tg}+C_{gg})C_{bg}V_{bg}/(eC_{gg})$ and $\mu = -eC_{bg}V_{bg}/C_{gg}$ vice-versa. Our overall results are shown in Fig. 3(c). The behavior of the experimental points is in agreement with the expected Dirac dispersion $\mu = \text{sign}(n)\hbar v_F(\pi|n|)^{1/2}$, with $v_F$ = 0.97×10$^6$ m/s (1.15 ×10$^6$ m/s) estimated for the lower (upper) layer. Since we are unaware of physical mechanisms inducing a different $v_F$ in the two layers, we attribute the slight difference to the experimental accuracy in the determination of the CNP position, which is limited by broadening of the CNP by charge fluctuations in each layer (~5×10$^9$ cm$^{-2}$).

By applying a moderate perpendicular magnetic field ($B$ = 250 mT, Fig. 3(d)), we observe a series of gate-tunable interlayer quantum Hall states [25]. The LLs from the two layers are bound to transitions between such states, giving large values of $d\sigma_{xy}/dV_{tg}$ (first derivative of the Hall conductivity) [23]. The experimental pattern in Fig. 3(d) is well reproduced by the computed LLs positions as a function of the gate potentials, reported in Fig. 3(e), where $n_{upper(lower)} = eB/h \times 4N$ ($N$ = 0, ±1, ±2, … and 4 accounts for the spin and valley degeneracy in each layer). However, we note that this calculation does not consider the quantization of the energy spectrum, and therefore cannot quantitatively match all the details in the experimental data. In presence of $B$, the thermodynamic measurements can rely on the crossings between the zero-energy LL of the probe layer (pinned at the probe CNP) and the finite-energy LLs of the target layer. As shown in the inset of Fig. 3(f), each of these crossings results in an 8-fold quantized step in $\sigma_{xy}$. Considering the position of the corresponding maximum in $d\sigma_{xy}/dV_{tg}$ (red arrow in Fig. 3(f) inset) and using the same relations as in zero field, we can obtain the chemical potential at the target LLs, shown in Fig. 3(f). Due to a general lower quality of the $d\sigma_{xy}/dV_{tg}$ data for $V_{bg} > 0$, here we employ only the crossings at $V_{bg} < 0$, providing $\mu$ for the upper (lower) layer for $N > 0$ ($N < 0$). We compare these data with the LL energies $E(N)$ = sign(N)$v_F$($2e\hbar B|N|)^{1/2}$, (light blue diamonds in Fig. 3(f), $v_F$ = $10^6$ m/s). The measurement correctly captures the $|N|^{1/2}$ dependence, and the large scale of $E(\pm 1)$ with respect to the energy separation of the higher-index LLs, both hallmarks of the Dirac dispersion. The standard deviation between the theoretical LL energies and the experimental $\mu$ values is 1.4 meV, which sets our current resolution for thermodynamic measurements in magnetic fields.

In principle, our approach can be used to obtain high-resolution thermodynamic information on an arbitrary graphene-based system (for instance, magic-angle TBG as in Ref.[20]), simply by stacking it on top of SLG and imposing a large relative twisting. We stress that the 30° rotation is not a strict requirement: smaller twist angles could be equivalently employed, as long as the interlayer decoupling is preserved [26,27]. While in the case presented here of SLG probing SLG, the value of the interlayer capacitance $C_{gg}$ is well established, this might vary significantly when considering a different target system. However, given the reciprocity of the technique, $C_{gg}$ can be obtained by measuring $\mu(n)$ of the probe SLG via tracking the target CNP, initializing a reasonable starting value of interlayer capacitance (e.g that of large-angle TBG) and iterating the procedure until convergence to the SLG Dirac dependence. We note that both the parallel transport mechanism and the thermodynamic measurement scheme could be extended to other van der Waals systems where interlayer decoupling is obtained from large-angle twisting, such as twisted transition metal dichalcogenides.

In summary, we have shown that it is possible to achieve simultaneous ultra-high mobility and conductivity in 30TBG, based on parallel conduction and gate-controlled carrier distribution. In addition, we have exploited the low-energy electronic decoupling of 30TBG to introduce a technique for layer-sensitive thermodynamic measurements using a standard device structure and measurement configuration.

**Figures:**

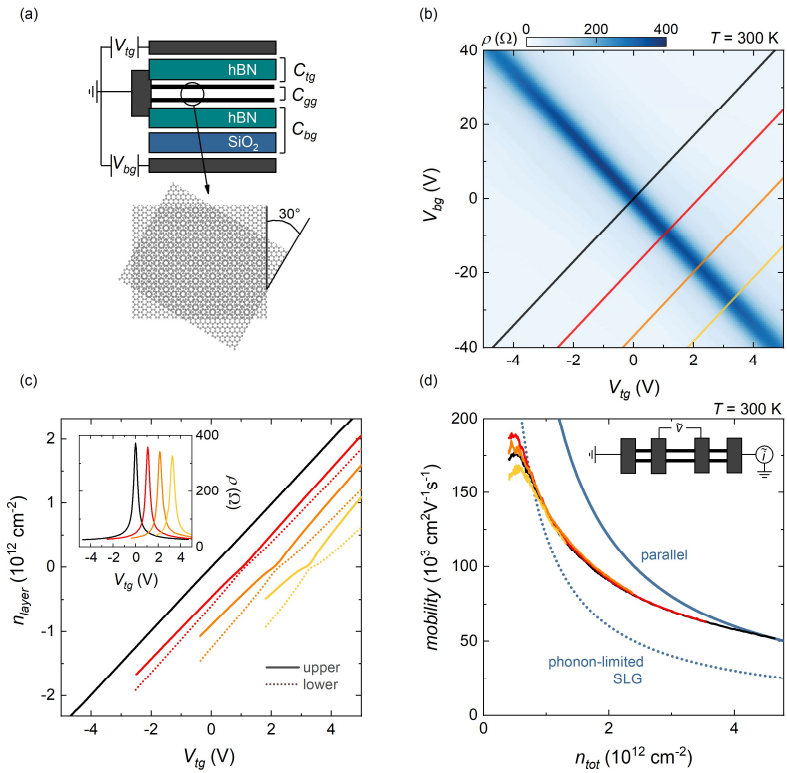

FIG. 1. Parallel transport and ultra-high mobility at room temperature. (a) Schematics of the lateral section of the investigated device. CVD-grown 30°-twisted bilayer graphene is encapsulated between hBN flakes. The potentials $V_{tg}$ and $V_{bg}$ are applied to the top and back-gate, respectively. The capacitance between the top-gate and the upper graphene layer ($C_{tg}$), between the two graphene layers ($C_{gg}$), and between the lower graphene layer and the back-gate ($C_{bg}$) are indicated. (b) Longitudinal resistivity of TBG as a function of $V_{tg}$ and $V_{bg}$, measured at $T$ = 300 K. The lines indicate different trajectories of the type $V_{tg} \times C_{tg} - V_{bg} \times C_{bg}$ = *const*, employed in panels (c) and (d). (c) Computed charge density of the upper (solid lines) and lower (dashed lines) graphene layer along the straight trajectories indicated in panel (b). Inset: longitudinal resistivity peaks as a function of $V_{tg}$ along the same lines. (d) Mobility of 30TBG as a function of $n_{tot}$ along the four trajectories. The dashed blue line is the intrinsic phonon-limited mobility of SLG [28], the solid blue line is the intrinsic mobility of two charge-balanced phonon-limited SLG conducting in parallel. Inset: sketch of the four-terminal measurement configuration adopted on the TBG device.

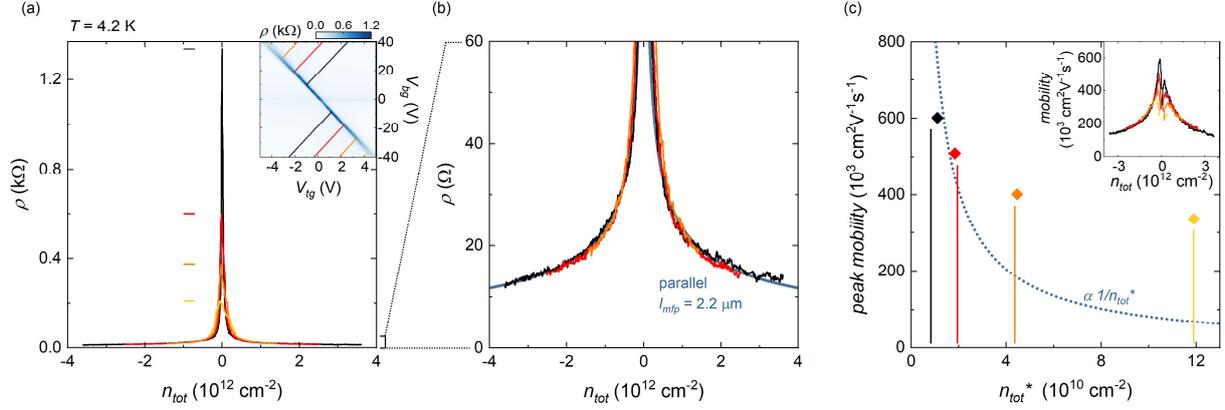

FIG. 2. Low-temperature parallel transport. (a) Longitudinal resistivity as a function of $n_{tot}$, measured at $T$ = 4.2 K along the straight trajectories indicated by the lines in the inset. The horizontal markers indicate the maxima of the resistivity peaks. Inset: low $T$ resistivity of 30TBG as a function of $V_{tg}$ and $V_{bg}$. (b) Zoom-in of the longitudinal resistivity as a function of $n_{tot}$. The blue line is the calculated resistivity of two charge-balanced SLG conducting in parallel, with a carrier mean free path ($l_{mfp}$) equal to the width of our device channel (2.2 μm). (c) Measured peak mobility along the different trajectories (maximum values in the inset) as a function of $n_{tot}^*$ (diamonds). $n_{tot}^*$ values obtained from parallel transport simulations (assuming $l_{mfp}$ = 2.2 μm) are indicated by the vertical lines. The dashed blue line shows the $n_{tot}^*$-dependence of the mobility from Ref. [40], considering parallel transport among two charge-balanced SLG. Inset: mobility of 30TBG as a function of $n_{tot}$ along the trajectories in panel (a) inset.

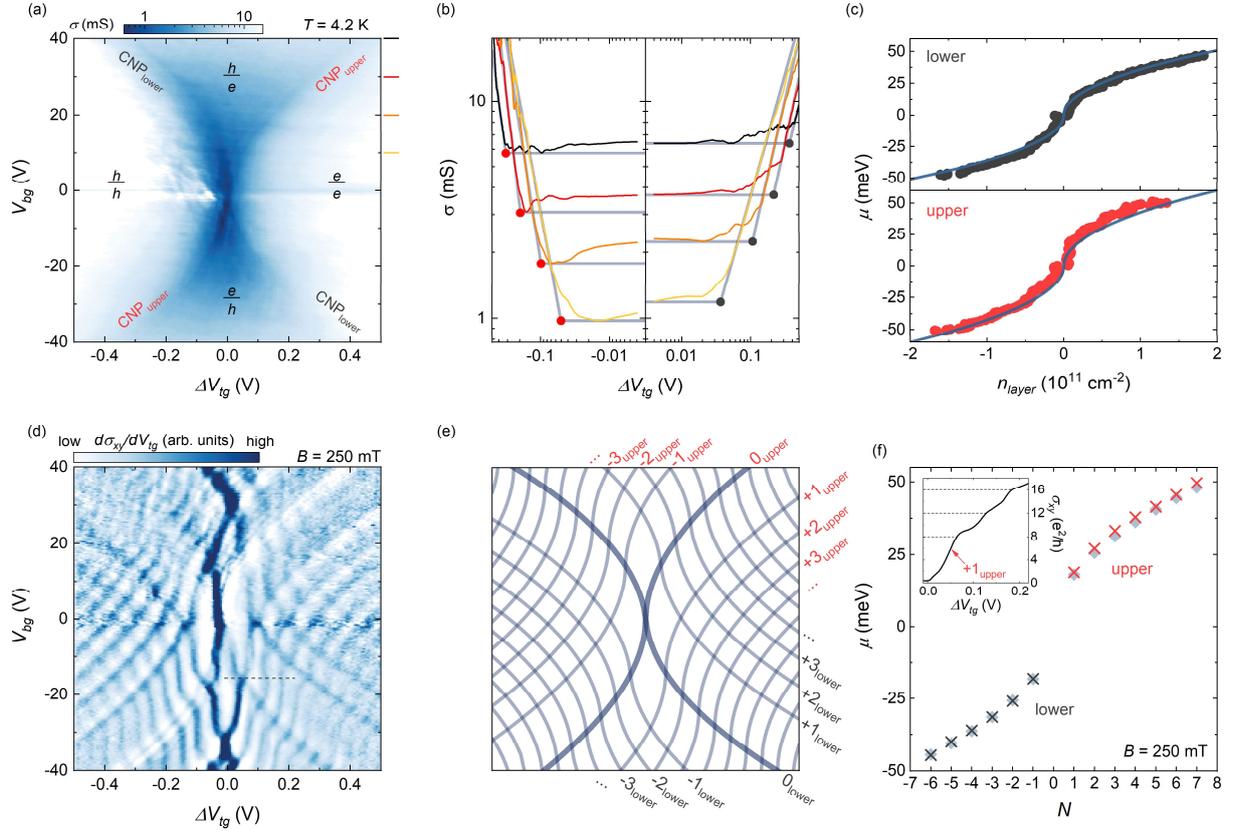

FIG. 3. Layer-resolved thermodynamic measurements in 30TBG. (a) Longitudinal conductivity as a function of $\Delta V_{tg}$ (top-gate voltage relative to the global CNP) and $V_{bg}$. The hourglass-like shape in the color plot is due to the split CNPs for the two layers, leading to the four interlayer charge configurations indicated. (b) Log($\sigma$) as a function of log($\Delta V_{tg}$) along horizontal cuts from panel (a) ($V_{bg}$ = 40 V to 10 V, see markers in panel (a)). In the left (right) part, linear fits to Log($\sigma$) intersect the minimum conductivity at the CNP of the lower (upper) layer, acting as probe. The corresponding values of gate voltages at the red (black) intersecting points are used to extract $\mu$ and $n$ in the upper (lower) layer, acting as target. (c) Experimentally measured chemical potential as a function of the carrier density for the two graphene layers (black and red circles). The blue lines are fits to the Dirac dispersion, giving the $v_F$ values reported in the text. (d) First derivative of the Hall conductivity, as a function of $\Delta V_{tg}$ and $V_{bg}$, measured at $B$ = 250 mT and $T$ = 4.2 K. (e) Theoretical LLs positions for the two layers as a function of $\Delta V_{tg}$ and $V_{bg}$ (same scales as in panel (d)), computed according to the electrostatic model described in Supplemental Material [22]. (f) Chemical potential of the upper (red crosses) and lower layer (black crosses) as a function of the LL index $N$. The light blue diamonds are the theoretical LL energies at $B$ = 250 mT. Inset: Hall conductivity as a function of $\Delta V_{tg}$, along the dashed line in panel (d). The horizontal lines indicate the quantized plateau values; slight deviations are attributed to residual bulk conduction at low magnetic field. The arrow indicates the crossing point between the lower $N$ = 0 level and the upper $N$ = +1, individuated at the maximum $d\sigma_{xy}/dV_{tg}$.

**Note added:** Most recent work on double-gated 30°-twisted [43] and large-angle-twisted [44] bilayer graphene show low-temperature transport features consistent with our findings.

**Acknowledgments:** We thank F. Rossella for technical support during the low temperature experiments. Growth of hexagonal boron nitride crystals was supported by the Elemental Strategy Initiative conducted by the MEXT, Japan, Grant Number JPMXP0112101001, JSPS KAKENHI Grant Numbers JP20H00354 and the CREST(JPMJCR15F3), JST. The research leading to these results has received funding from the European Union's Horizon 2020 research and innovation program under grant agreements no. 785219-Graphene Core2 and 881603-Graphene Core3.

*To whom correspondence should be addressed: camilla.coletti@iit.it

**To whom correspondence should be addressed: sergio.pezzini@nano.cnr.it

**References:**

[1] Y. M. Lin, K. A. Jenkins, V. G. Alberto, J. P. Small, D. B. Farmer, and P. Avouris, Operation of graphene transistors at gigahertz frequencies, Nano Lett. **9**, 422 (2009).

[2] C. H. Yeh, P. Y. Teng, Y. C. Chiu, W. T. Hsiao, S. S. H. Hsu, and P. W. Chiu, Gigahertz Field-Effect Transistors with CMOS-Compatible Transfer-Free Graphene, ACS Appl. Mater. Interfaces **11**, 6336 (2019).

[3] V. Sorianello, M. Midrio, G. Contestabile, I. Asselberghs, J. Van Campenhout, C. Huyghebaert, I. Goykhman, A. Ott, A. Ferrari, and M. Romagnoli, Graphene-silicon phase modulators with gigahertz bandwidth, Nat. Photonics **12**, 40 (2018).

[4] A. V. Rozhkov, A. O. Sboychakov, A. L. Rakhmanov, and F. Nori, Electronic properties of graphene-based bilayer systems, Phys. Rep., **648** (2016).

[5] J. Hass J. E. Millán-Otoya, M. Sprinkle, N. Sharma, W. A. de Heer, C. Berger, P. N. First, L. Magaud, and E. H. Conrad, Why multilayer graphene on 4H-SiC(000$\bar{1}$) behaves like a single sheet of graphene, Phys. Rev. Lett. **100**, 125504 (2008).

[6] R. Negishi, Y. Ohno, K. Maehashi, K. Matsumoto, and Y. Kobayashi, Carrier transport properties of the field effect transistors with graphene channel prepared by chemical vapor deposition, Jpn. J. Appl. Phys. **51**, 06FD03 (2012).

[7] K. Uemura, T. Ikuta, and K. Maehashi, Turbostratic stacked CVD graphene for high-performance devices, Jpn. J. Appl. Phys. **57**, 030311 (2018).

[8]  X. Wu, Y. Chuang, A. Contino, B. Sorée, S. Brems, Z. Tokei, M. Heyns, C. Huyghebaert, and I. Asselberghs, Boosting Carrier Mobility of Synthetic Few Layer Graphene on SiO2 by Interlayer Rotation and Decoupling, Adv. Mater. Interfaces **5**, 1800454 (2018).

[9]  M. S. Kim, M. Kim, S. Son, S.-Y. Cho, S. Lee, D.-K. Won, J. Ryu, I. Baes, H.-M. Kim, and K.-B. Kim, Sheet Resistance Analysis of Interface-Engineered Multilayer Graphene: Mobility Versus Sheet Carrier Concentration, ACS Appl. Mater. Interfaces **12**, 30932 (2020).

[10] S. Shallcross, S. Sharma, E. Kandelaki, and O. A. Pankratov, Electronic structure of turbostratic graphene, Phys. Rev. B **81**, 165105 (2010).

[11] Y. Gao and S. Okada, Carrier distribution control in bilayer graphene under a perpendicular electric field by interlayer stacking arrangements, Appl. Phys. Express **14**, 035001 (2021).

[12] C. P. Lu, M. Rodriguez-Vega, G. Li, A. Luican-Mayer, K. Watanabe, T. Taniguchi, E. Rossi, and E. Y. Andrei, Local, global, and nonlinear screening in twisted double-layer graphene, Proc. Natl. Acad. Sci. U. S. A. **113**, 6623 (2016).

[13] J. Martin, N. Akerman, G. Ulbricht, T. Lohmann, J. Smet, K. Klitzing, and A. Yacoby, Observation of electron-hole puddles in graphene using a scanning single-electron transistor, Nat. Phys. **4**, 144 (2008).

[14] J. Martin, B. E. Feldman, R. T. Weitz, M. T. Allen, and A. Yacoby, Local compressibility measurements of correlated states in suspended bilayer graphene, Phys. Rev. Lett. **105**, 256806 (2010).

[15] A. F. Young, C. R. Dean, I. Meric, S. Sorgenfrei, H. Ren, K. Watanabe, T. Taniguchi, J. Hone, K. L. Shepard, and P. Kim, Electronic compressibility of layer-polarized bilayer graphene, Phys. Rev. B **85**, 235458 (2012).

[16] E. A. Henriksen and J. P. Eisenstein, Measurement of the electronic compressibility of bilayer graphene Phys. Rev. B **82**, 041412(R) (2010).

[17] A. A. Zibrov, C. Kometter, H. Zhou, E. M. Spanton, T. Taniguchi, K. Watanabe, M. P. Zaletel, and A. F. Young, Tunable interacting composite fermion phases in a half-filled bilayer-graphene Landau level, Nature **549**, 360 (2017).

[18] S. Kim Insun Jo, D. C. Dillen, D. A. Ferrer, B. Fallahazad, Z. Yao, S. K. Banerjee, and E. Tutuc, Direct measurement of the fermi energy in graphene using a double-layer heterostructure, Phys. Rev. Lett. **108**, 116404 (2012).

[19] K. Lee, B. Fallahazad, J. Xue, D. C. Dillen, K. Kim, T. Taniguchi, K. Watanabe, and E. Tutuc, Chemical potential and quantum Hall ferromagnetism in bilayer graphene, Science **345**, 6192 (2014).

[20] J. M. Park, Y. Cao, K. Watanabe, T. Taniguchi, and P. Jarillo-Herrero, Flavour hund's coupling, correlated chern gaps, and diffusivity in moiré flat bands, Nature **592**, 43 (2021).

[21] Fangyuan Yang, Alexander A. Zibrov, Ruiheng Bai, Takashi Taniguchi, Kenji Watanabe, Michael P. Zaletel, and Andrea F. Young, Experimental Determination of the Energy per Particle in Partially Filled Landau Levels, Phys. Rev. Lett. **126**, 156802 (2021).

[22] See Supplemental Material at [URL will be inserted by publisher] for details on device fabrication and measurement setup, description of the electrostatic model, additional low-temperature resistivity curves, discussion on extrinsic disorder in CVD-grown 30TBG, additional data on *e-h* coexistence at high displacement field, discussion on gate-dependent total carrier density in the *e-h* configuration.

[23] S. Pezzini, V. Miseikis, G. Piccinini, S. Forti, S. Pace, R. Engelke, F. Rossella, K. Watanabe, T. Taniguchi, P. Kim, C. Coletti, 30°-Twisted Bilayer Graphene Quasicrystals From Chemical Vapor Deposition, Nano Lett. **20**, 3313 (2020).

[24] Z. Yan, Y. Liu, Long Ju, Z. Peng, J. Lin, G. Wang, H. Zhou, C. Xiang, E. L. G. Samuel, C. Kittrell, V. I. Artyukhov, F. Wang, B. I. Yakobson, J. M. Tour, Large Hexagonal Bi-and Trilayer Graphene Single Crystals with Varied Interlayer Rotations, Angew. Chem. **53**, 1591 (2014).

[25] D. G. Purdie, N. M. Pugno, T. Taniguchi, K. Watanabe, A. C. Ferrari, and A. Lombardo, Cleaning interfaces in layered materials heterostructures, Nat. Commun. **9**, 5387 (2018).

[26] J. D. Sanchez-Yamagishi, T. Taychatanapat, K. Watanabe, T. Taniguchi, A. Yacoby, and P. Jarillo-Herrero, Quantum hall effect, screening, and layer-polarized insulating states in twisted bilayer graphene, Phys. Rev. Lett. **108**, 076601 (2012).

[27] P. Rickhaus, M.-H. Liu, M. Kurpas, A. Kurzmann, Y. Lee, H. Overweg, M. Eich, R. Pisoni, T. Taniguchi, K. Watanabe, K. Richter K. Ensslin, and T. Ihn, The electronic thickness of graphene, Sci. Adv. **6**, eaay8409 (2020).

[28] E. H. Hwang and S. Das Sarma, Acoustic phonon scattering limited carrier mobility in two-dimensional extrinsic graphene, Phys. Rev. B **77**, 115449 (2008).

[29] C. H. Park, N. Bonini, T. Sohier, G. Samsonidze, B. Kozinsky, M. Calandra, F. Mauri, and N. Marzari, Electron-phonon interactions and the intrinsic electrical resistivity of graphene, Nano Lett. **14**, 1113 (2014).

[30] L. Wang, I. Meric, P. Y. Huang, Q. Gao, Y. Gao, H. Tran, T. Taniguchi, K. Watanabe, L. M. Campos, D. A. Muller, J. Guo, P. Kim, J. Hone, K. L. Shepard, C. R. Dean, One-dimensional electrical contact to a two-dimensional material, Science **342**, 614 (2013).

[31] Y. M. Lin, C. Dimitrakopoulos, K. A. Jenkins, D. B. Farmer, H.-Y. Chiu, A. Grill, and Ph. Avouris, 100-GHz transistors from wafer-scale epitaxial graphene, Science **327**, 662 (2010).

[32] Y. Lin, H. Chiu, K. A. Jenkins, D. B. Farmer, P. Avouris, and A. Valdes-Garcia, Dual-Gate Graphene FETs With $f_{T}$ of 50 GHz, IEEE Electron Device Lett. **31**, 68 (2009).

[33] L. Viti, D. G. Purdie, A. Lombardo, A. C. Ferrari, and M. S. Vitiello, HBN-Encapsulated, Graphene-based, Room-temperature Terahertz Receivers, with High Speed and Low Noise, Nano Lett. **20**, 3169 (2020).

[34] F. Bonaccorso, Z. Sun, T. Hasan, and A. C. Ferrari, Graphene photonics and optoelectronics, Nat. Photonics **4**, 611 (2010).

[35] M. A. Giambra, V. Mišeikis, S. Pezzini, S. Marconi, A. Montanaro, F. Fabbri, V. Sorianello, A. C. Ferrari, C. Coletti, and M. Romagnoli, Wafer-Scale Integration of Graphene-Based Photonic Devices, ACS Nano **15**, 3171 (2021).

[36] Y. Wang, S. W. Tong, X. F. Xu, B. Özyilmaz, and K. P. Loh, Interface engineering of layer-by-layer stacked graphene anodes for high-performance organic solar cells, Adv. Mater. **23**, 1514 (2011).

[37] X. Kong, L. Zhang, B. Liu, H. Gao, Y. Zhang, H. Yan, and X. Song, Graphene/Si Schottky solar cells: a review of recent advances and prospects, RSC Adv. **9**, 863 (2019).

[38] Y. Ohno, K. Maehashi, Y. Yamashiro, and K. Matsumoto, Electrolyte-Gated Graphene Field-Effect Transistors for Detecting pH and Protein Adsorption, Nano Lett. **9**, 3318 (2009).

[39] J. Dauber, A. A. Sagade, M. Oellers, K. Watanabe, T. Taniguchi, D. Neumaier, and C. Stampfer, Ultra-sensitive Hall sensors based on graphene encapsulated in hexagonal boron nitride Appl. Phys. Lett. **106**, 193501 (2015)

[40] N. J. G. Couto, D. Costanzo, S. Engels, D.-K. Ki, K. Watanabe, T. Taniguchi, C. Stampfer, F. Guinea, and A. F. Morpurgo, Random strain fluctuations as dominant disorder source for high-quality on-substrate graphene devices, Phys. Rev. X **4**, 041019 (2014).

[41] Y. Kim, H. Yun, S.-G. Nam, M. Son, D. S. Lee, D. C. Kim, S. Seo, H. C. Choi, H.-J. Lee, S. W. Lee, and J. S. Kim, Breakdown of the interlayer coherence in twisted bilayer graphene, Phys. Rev. Lett. **110**, 096602 (2013).

[42] R. Bistritzer and A. H. MacDonald, Transport between twisted graphene layers, Phys. Rev. B **81**,

245412 (2010).

[43] S. Slizovskiy, A. Garcia-Ruiz, A. I. Berdyugin, N. Xin, T. Taniguchi, K. Watanabe, A. K. Geim, N. D. Drummond, and V. I. Fal'ko, Out-of-Plane Dielectric Susceptibility of Graphene in Twistronic and Bernal Bilayers, Nano Lett. **21**, 6678 (2021).

[44] A. Mrenca-Kolasinska, P. Rickhaus, G. Zheng, K. Richter, T. Ihn, K. Ensslin, and M-H. Liu, Quantum capacitive coupling between large-angle twisted graphene layers, arXiv:2110.00907 (2021).

*Supplemental Material for*

# Parallel transport and layer-resolved thermodynamic measurements in twisted bilayer graphene

G. Piccinini,[1,2] V. Mišeikis,[2,3] K. Watanabe,[4] T. Taniguchi,[5] C. Coletti[2,3,*], S. Pezzini[6,**]

[1]*NEST, Scuola Normale Superiore, Piazza San Silvestro 12, 56127 Pisa, Italy*
[2]*Center for Nanotechnology Innovation @NEST, Istituto Italiano di Tecnologia, Piazza San Silvestro 12, 56127 Pisa, Italy*
[3]*Graphene Labs, Istituto Italiano di Tecnologia, Via Morego 30, 16163 Genova, Italy*
[4]*Research Center for Functional Materials, National Institute for Materials Science, 1-1 Namiki, Tsukuba, 305-0044, Japan*
[5]*International Center for Materials Nanoarchitectonics, National Institute for Materials Science, 1-1 Namiki, Tsukuba, 305-0044, Japan*
[6]*NEST, Istituto Nanoscienze-CNR and Scuola Normale Superiore, Piazza San Silvestro 12, Pisa 56127, Italy*

**Device fabrication and measurement setup**

The Hall bar device is fabricated on encapsulated 30TBG using standard e-beam lithography (EBL), combined with reactive ion etching and metal evaporation. First, edge contacts are fabricated by EBL patterning of a polymer mask (PMMA 950 A4), full etching of the heterostructure in $CF_4/O_2$ and thermal evaporation of Cr/Au (5/50 nm). The RIE recipe consists in a flow of 20 sccm of $CF_4$ and 2 sccm of $O_2$, with a power of 25 W, a pressure of 0.06 mbar, maintained for 30 seconds. The single-step procedure minimizes contamination of the TBG edges prior to metallization. Secondly, a top-gate (Cr/Au 5/50 nm) is fabricated, avoiding short circuit with the electrical contacts and with exposed TBG edges. Finally, the heterostructure is etched (same recipe as above) in a 6-terminal Hall bar shape, using the physical mask provided by the metallic top gate, in combination with additional e-beam-patterned PMMA arms. The Hall bar is 2.2 μm wide, the distance between the voltage probes is 3 μm.

The sample is mounted in a variable-temperature insert, loaded into a liquid-helium cryostat with superconducting coil. All the measurements are performed in four-probe configuration, with constant current excitation (10-100 nA) and standard lock-in acquisition (13 Hz).

**Electrostatic model**

Modelling the electrostatics of dual-gated TBG allows obtaining the individual carrier density in the two layers ($n_{upper}$ and $n_{lower}$) at arbitrary values of the applied gate voltages ($V_{tg}$ and $V_{bg}$). Sanchez-Yamagishi *et al.* [1] firstly reported on the screening properties of large-angle TBG, and pointed out the importance of including the chemical potential of the individual layers in the electrostatic description (see SM file therein [1]). Equivalent approaches, although with different naming of the relevant quantities, were used more recently by Park *et al.* [2] (hBN-spaced double-layer), and Rickhaus *et al.* [3] (large-angle TBG, with and without hBN spacer). Following the definitions of Ref. [1], the total carrier density and electric displacement field are given by

$$n_{tot} = [C_{tg}(V_{tg} - \mu_{upper}/e) + C_{bg}(V_{bg} - \mu_{lower}/e)]/e \qquad (S1)$$

$$D = [C_{tg}(V_{tg} - \mu_{upper}/e) - C_{bg}(V_{bg} - \mu_{lower}/e)]/2 \qquad (S2)$$

where $V_{tg}$ and $V_{bg}$ are the applied gate voltages, and the chemical potential $\mu_{upper(lower)}$ is connected to the carrier density in the layer via the standard relation for Dirac fermions $\mu_{upper(lower)} = sign(n_{upper(lower)}) \hbar v_F \sqrt{\pi |n_{upper(lower)}|}$. The applied top-gate (back-gate) voltages are partially screened (i.e. reduced) by the chemical potential in the upper (lower) graphene layer. Such screening effect is central in the electrostatic description of large-angle TBG, and needs to be included both in the total carrier density and displacement electric field. The total carrier density is then determined by the sum of the screened potentials, the displacement field by their difference. The top- and back-gate capacitance are known from the thickness and dielectric constant of the gate dielectrics (see main text). The electric displacement field $D$ is screened by both a charge imbalance between the layers $\Delta n = n_{upper} - n_{lower}$ and the interlayer dielectric environment (interlayer capacitance $C_{gg}$, value taken from Ref. [3]), which determines a chemical potential difference $\Delta\mu = \mu_{upper} - \mu_{lower}$ [1]. Therefore, the displacement field in Equation S2 can be split into two components, according to

$$D = C_{gg} \frac{\Delta\mu}{e} + e \frac{\Delta n}{2} = D_{\Delta\mu} + D_{\Delta n} \qquad (S3)$$

To determine $n_{upper}$ as a function of $V_{tg}$ and $V_{bg}$, we use the following procedure (equivalently for $n_{lower}$):

1. Take a dense sequence of equally-spaced values of $n_{tot}$ over an experimentally relevant range, e.g. [-4.8×10$^{12}$; 4.8×10$^{12}$] cm$^{-2}$.
2. Fix $n_{upper}$ to an arbitrary value -2.4×10$^{12}$ cm$^{-2}$ < $n_{upper}$ < 2.4×10$^{12}$ cm$^{-2}$
3. For each $n_{tot}$, calculate the corresponding $n_{lower} = n_{tot} - n_{upper}$, $\Delta n$ and $D_{\Delta n}$
4. For each $n_{tot}$, calculate $\mu_{lower}$ from $n_{lower}$ ($\mu_{upper}$ is fixed by $n_{upper}$), $\Delta\mu$ and $D_{\Delta\mu}$
5. For each $n_{tot}$, calculate $D$
6. Use the inverted expressions of Equation S1 and S2 to calculate $V_{tg}$ and $V_{bg}$ from the quantities obtained at the previous steps, defining points of constant $n_{upper}$

$$V_{tg} = (-2\epsilon_0 \frac{D}{e} + n_{tot} + 2\frac{C_{tg}}{e}\mu_{upper})/(2\frac{C_{tg}}{e})$$

$$V_{bg} = (-2\epsilon_0 \frac{D}{e} + n_{tot} + 2\frac{C_{bg}}{e}\mu_{lower})/(2\frac{C_{bg}}{e})$$

7. Go back to step 2 and iterate the procedure for arbitrary values of $n_{upper}$.

Our complete results are shown in Figure S1(a) ($n_{upper}$) and Figure S1(b) ($n_{lower}$). The same procedure can be repeated to obtain the layers' carrier density in the zoomed-in $\Delta V_{tg}$-$V_{bg}$ space (Figure S1(c) and S1(d)). Based on these plots, we can express the experimental data (obtained as a function of the gate potentials along a given $V_{tg}$-$V_{bg}$ trajectory) as a function of the total carrier density (see main text Figure 1 and Figure 2). We use an algorithm that calculates the voltage distances between each point of the chosen experimental trajectory and all the points in both Figures S1(a) and (b), and then finds the minimum distances. Next, the algorithm associates to each point of the experimental trajectory a total charge density given by the sum of $n_{upper}$ and $n_{lower}$ corresponding to the minimum distance point.

**Low temperature resistivity curves**

The low $T$ resistivity curves in main text Figure 2(a) are plotted considering half of the $\rho$ data from the lower left quadrant ($n_{tot} < 0$) of the $V_{tg}$-$V_{bg}$ space, half from the upper right one ($n_{tot} > 0$). This is done to avoid a small resistive peak independent from $V_{tg}$, and located at $V_{bg} \sim 0$ V, shown in Figure S2. The insensitivity to $V_{tg}$ indicates that this feature originates from the contact areas, which are not covered by the top-gate. At $V_{bg} \sim 0$ V, the TBG close to the contacts is undoped and minor malfunctioning is likely.

**Extrinsic disorder in CVD-grown 30TBG**

The use CVD-grown 30TBG crystals is mainly motivated by the capability of avoiding the van der Waals assembly step and obtaining TBG with high yield (tens of samples from a single growth cycle) [4]. However, in analogy to recent reports on CVD-grown SLG [5,6], such choice does not harm the quality of the electrical transport when combined with dry hBN encapsulation. The following observations from low $T$ experiments support this claim:

(i) from the width of the CNP region ($n^*_{tot}$) at low displacement field, the charge fluctuations in our device are $\sim 10^{10}$ cm$^{-2}$, which corresponds to $\sim 5\times 10^9$ cm$^{-2}$ in each graphene layer. Such magnitude is fully comparable with ultra-clean devices based on exfoliated flakes [7].

(ii) the carrier mean free path in each layer is $\sim 2.2$ μm (see Figure 2b), which equals the physical width of the device channel. Hence, the main scattering contribution can be identified with the sample edges, again matching the results of the community on exfoliation-based devices [8].

(iii) a complete set of interlayer quantum Hall states is resolved at $B$ = 250 mT (Figure 3d) and quantum oscillations start to be visible at ~50 mT (Figure S3 in SM), indicating low disorder broadening of the LLs [9].

**$e$-$h$ coexistence at high displacement field**

In order to prove the presence of charge carriers of opposite sign, we perform magnetotransport at high displacement field. In Figure S3 we show the derivative of the Hall conductivity as a function of $\Delta V_{tg}$ and the magnetic field (0 T < $B$ < 1 T), taken at $V_{bg}$ = -40 V. Two separated Landau fans develop from the charge neutrality points of the two layers. In the central region, electrons in the upper graphene layer, giving right-dispersing features, coexist with holes in the lower graphene layer, giving left-dispersing features.

**Gate-dependent total carrier density in the $e$-$h$ configuration**

In main text Figure 3(a), the low $T$ conductivity shows regions of different magnitude corresponding to different charge configurations, separated by the CNPs of the two layers. The conductivity in the $e$-$h$ (and $h$-$e$) region is found to be lower with respect to the case of equal carrier sign. Despite the fact that both layers have a finite carrier density, unless at $\Delta V_{tg}$ = $V_{bg}$ = 0, the absolute value of the total carrier density $|n_{tot}|=|n_{upper}|+|n_{lower}|$ keeps relatively low in the $e$-$h$ region, therefore limiting the measured TBG conductivity. From the maps in Figure S1(c) and S1(d), we can obtain $|n_{tot}|$ along different trajectories in the $\Delta V_{tg}$-$V_{bg}$ space. In Figure S4 we show that an "horizontal" movement along $V_{bg}$ = 0 (red points, $e$-$e$ ) yields a strong linear increase in $|n_{tot}|$, leading to large values of $\sigma$. Instead, a "vertical" movement along $\Delta V_{tg}$= 0 (black points, $h$-$e$) induces less $|n_{tot}|$, with a pronounced sublinear dependence, responsible for the lower $\sigma$ measured experimentally.

**Supplemental Figures**

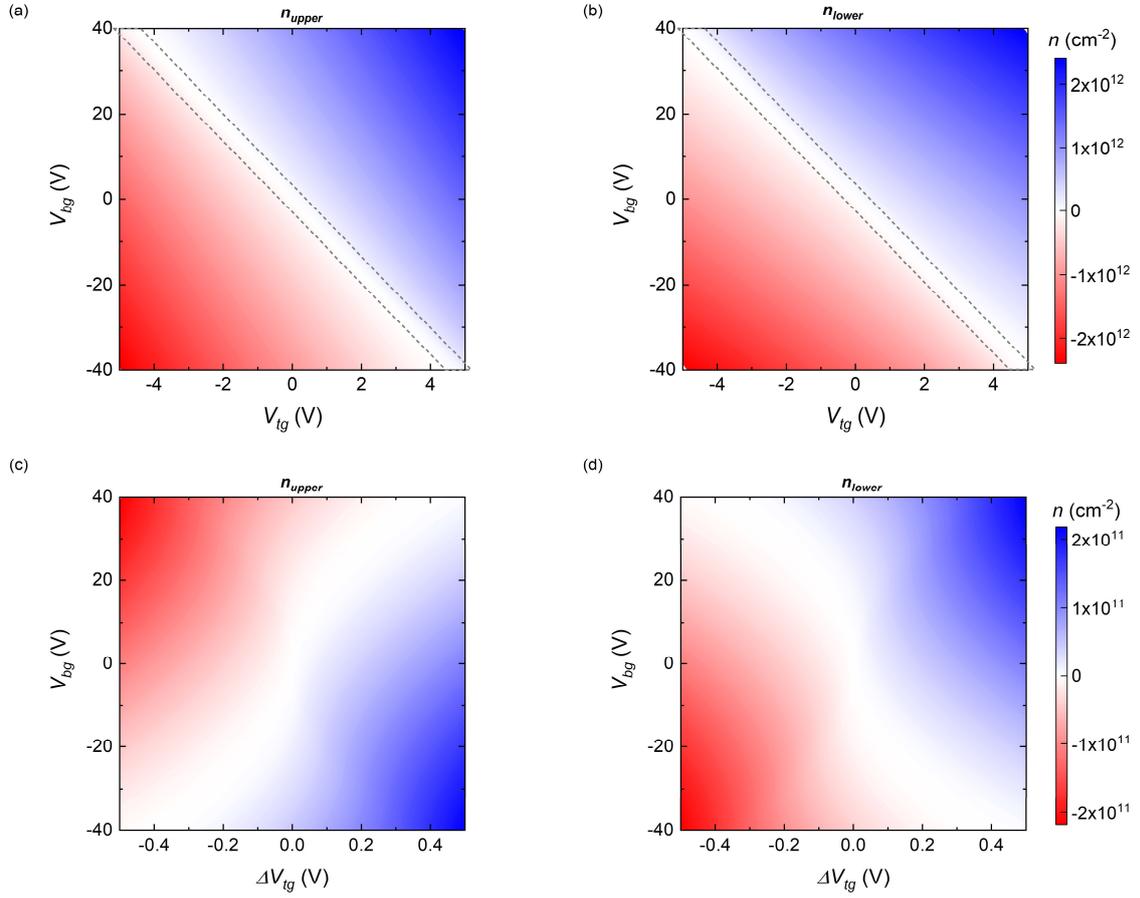

FIG. S1. (a, b) Computed charge density of the upper (a) and lower (b) graphene layers as a function of $V_{tg}$ and $V_{bg}$. (c, d) Computed charge density of the upper (c) and lower (d) graphene layers in the $\Delta V_{tg}$-$V_{bg}$ space, (region indicated by the dashed lines in panels (a) and (b)).

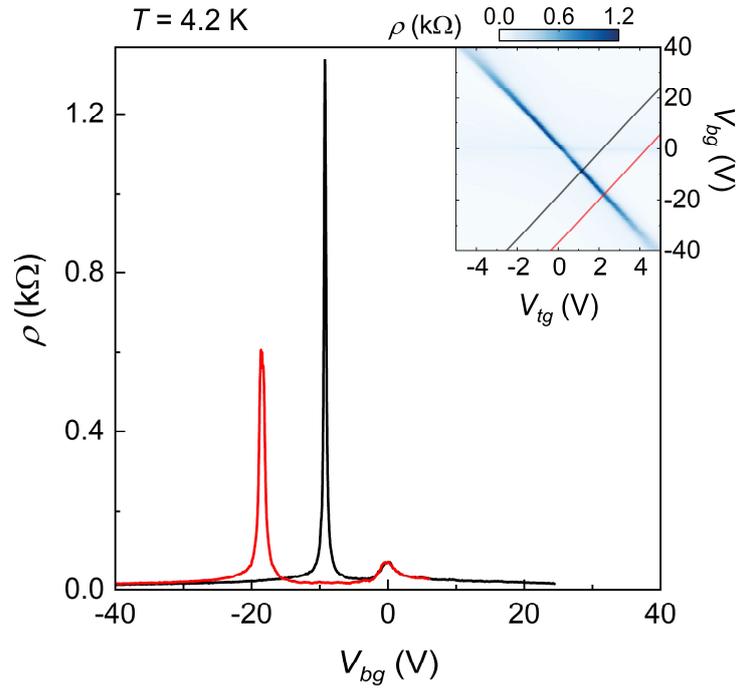

FIG. S2. Low-*T* longitudinal resistivity as a function of the back-gate voltage. In addition to the global CNP peak, a minor peak centered at $V_{bg} = 0$ is visible, independently of the $V_{tg}$-$V_{bg}$ trajectory (two examples are shown).

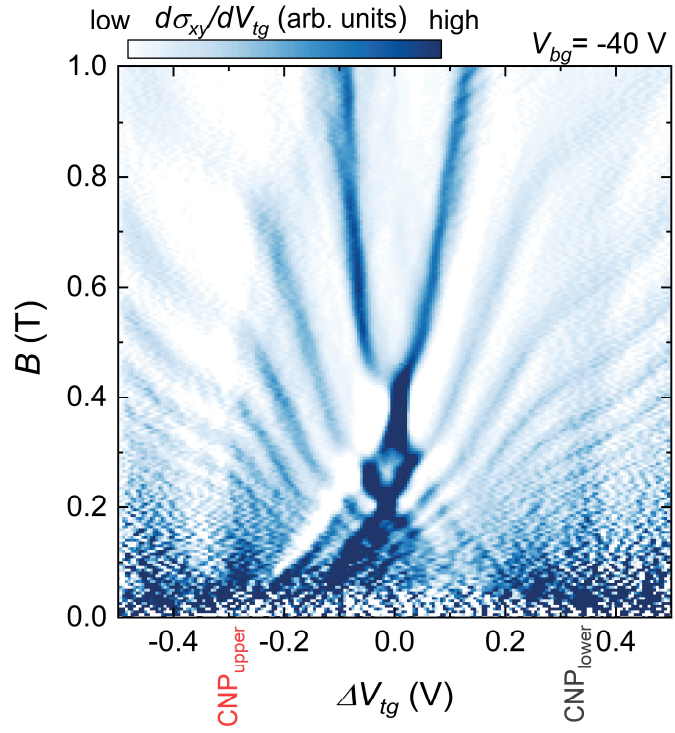

FIG. S3. First derivative of the Hall conductivity with respect to $V_{tg}$, as a function of $\Delta V_{tg}$ and $B$, measured at $V_{bg}$ = -40 V, $T$ = 4.2 K (the color scale is the same as in main text Figure 3(d)). Starting from $B \sim 50$ mT, two Landau fans with opposite dispersion cross in the area delimited by CNP$_{upper}$ and CNP$_{lower}$, demonstrating the coexistence of charge carriers of opposite sign.

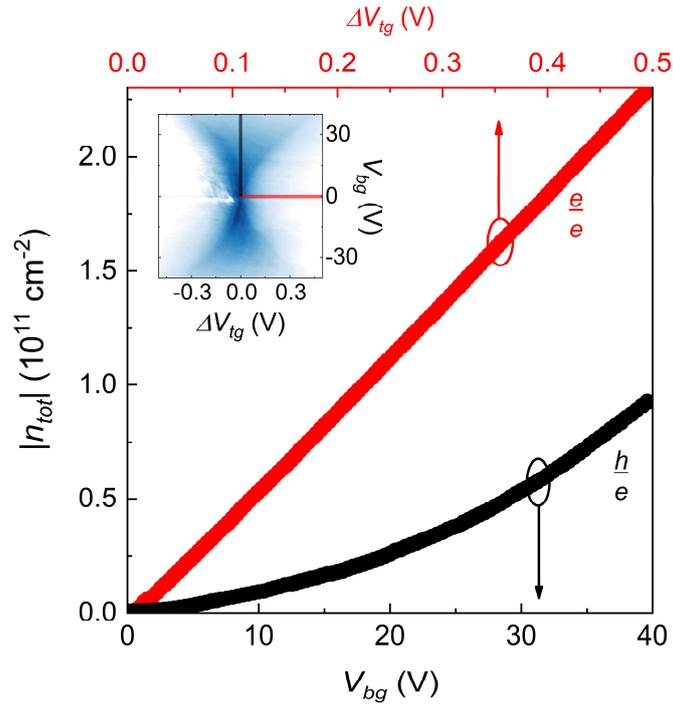

FIG. S4. Absolute value of the total carrier density along different trajectories in the $\Delta V_{tg}$-$V_{bg}$ space (shown by the red and black lines in the inset). The reduced carrier density in the *h-e* configuration determines the lower conductivity with respect to the *e-e* case (see color map in the inset, same as main text Figure 3(a)).

**References**

[1] J. D. Sanchez-Yamagishi, T. Taychatanapat, K. Watanabe, T. Taniguchi, A. Yacoby, and P. Jarillo-Herrero, "Quantum hall effect, screening, and layer-polarized insulating states in twisted bilayer graphene," Phys. Rev. Lett., vol. 108, no. 7, pp. 1–7, 2012.

[2] J. M. Park, Y. Cao, K. Watanabe, T. Taniguchi, and P. Jarillo-Herrero, "Flavour Hund's coupling, Chern gaps and charge diffusivity in moiré graphene," Nature, vol. 592, pages 43–48, 2021.

[3] P. Rickhaus et al., "The electronic thickness of graphene," Sci. Adv., vol. 6, no. 11, 2020.

[4] S. Pezzini et al., Nano Lett., "30°-Twisted Bilayer Graphene Quasicrystals from Chemical Vapor Deposition", vol. 20, no. 5, pp. 3313–3319, 2020.

[5] S. Pezzini et al., "High-quality electrical transport using scalable CVD graphene", 2D Mater. 7, 041003, 2020.

[6] M. Schmitz et al., "Fractional quantum Hall effect in CVD-grown graphene", 2D Mater. 7, 041007, 2020.

[7] D. Rhodes et al., "Disorder in van der Waals heterostructures of 2D materials", Nat. Mater. 18, 541–549, 2019.

[8] Wang et al., "One-Dimensional Electrical Contact to a Two-Dimensional Material", Science 342, 614–617, 2013.

[9] Y. Zeng et al., "High-Quality Magnetotransport in Graphene Using the Edge-Free Corbino Geometry", Phys. Rev. Lett. 122, 137701, 2019.